\documentclass[twocolumn,prb]{revtex4}
\usepackage{graphicx,psfrag}
\begin{document}
\title{R\'{e}nyi entropy, mutual information, and fluctuation properties of Fermi liquids}
\author{Brian Swingle}
\email{bswingle@mit.edu}
\affiliation{Department of Physics, Massachusetts Institute of Technology, Cambridge, Massachusetts 02139}
\begin{abstract}
I compute the leading contribution to the ground state R\'{e}nyi entropy $S_{\alpha}$ for a region of linear size $L$ in a Fermi liquid.  The result contains a universal boundary law violating term simply related the more familiar entanglement entropy.  I also obtain a universal crossover function that smoothly interpolates between the zero temperature result and the ordinary thermal R\'{e}nyi entropy of a Fermi liquid.  Formulas for the entanglement entropy of more complicated regions, including non-convex and disconnected regions, are obtained from the conformal field theory formulation of Fermi surface dynamics.  These results permit an evaluation of the quantum mutual information between different regions in a Fermi liquid.  I also study the number fluctuations in a Fermi liquid.  Taken together, these results give a reasonably complete characterization of the low energy quantum information content of Fermi liquids.
\end{abstract}
\maketitle

\section{Introduction} Recently, motivated by studies of entanglement in quantum systems, there has been a profitable exchange of ideas between the quantum information and condensed matter communities.  Much of this exchange has focused on the physics of entanglement entropy, defined as the von Neumann entropy of the reduced density matrix of a subsystem.  Entanglement entropy is the basic quantity that provides a characterization of the entanglement properties of many body quantum states \cite{geo_ent,arealaw1}.  Insights gained from the study of entanglement led to a new class of variational many body states collectively known as tensor network states \cite{mera,peps,terg}.  In addition, entanglement entropy can characterize interesting topological and critical phases of matter.  Conformal field theories in $1+1$ dimensions are partially characterized by the leading universal term in the their entanglement entropy \cite{geo_ent,eeqft}.  On the other hand, most phases in more than one spatial dimension have an entanglement entropy that scales as the boundary of the spatial region considered \cite{arealaw1,holoee}.  This boundary term is non-universal and typically forces us to study sub-leading terms in the entanglement entropy to find universal quantities characteristic of a particular phase.

There exists a class of systems in higher dimensions that violate this boundary law for the entanglement entropy.  This class includes free fermions \cite{fermion1,fermion2,fermion3,fermion4,fermion5,widom_proof,bgs_ferm1}, Fermi liquids \cite{bgs_ferm2}, Weyl fermions in a magnetic field \cite{bgs_highe}, and more exotic systems including critical Fermi surfaces and Bose metals.  The apparent unifying theme is the existence of many gapless one dimensional degrees of freedom in all these systems.  Fermi liquids with a codimension one Fermi surface have long been known to be equivalent to a set of nearly decoupled one dimensional gapless modes \cite{fermion_rg1,fermion_rg2,fermion_rg3}.  The system of free Weyl fermions in a background magnetic field is also explicitly $1+1$ dimensional due to the presence of zero modes propagating along the magnetic field lines.  Even critical Fermi surfaces behave thermodynamically like a collection of $1+1$ dimensional critical fields with critical exponent $z_f \neq 1$ \cite{crit_fs}.  In all these cases we may think of one dimensional theories patched together to form a higher dimensional theory, but this picture is most concrete for the case of Fermi liquids.

In this paper I use the one dimensional framework to compute the R\'{e}nyi entropy, defined below, for interacting fermions in a Fermi liquid state.  This quantity is interesting because it gives in principle complete knowledge of the spectrum of the reduced density matrix, although I will only compute it in the low energy limit.  I also describe the form of the entanglement entropy of more general regions (non-convex or even disjoint) using the same one dimensional formulation.  This permits a calculation of the mutual information between two distant regions in a Fermi liquid.  All of these information theoretic quantities turn out to provide direct access to the geometry of the interacting Fermi surface.  They are universal in the sense that they depend only on the geometry of the interacting Fermi surface and not on any other details of the Fermi liquid fixed point.  Additionally, I compute the fermion number fluctuations in a region of size $L$ in a Fermi liquid and give general finite temperature crossover forms for both the entropy and the number fluctuations. Together these results provide a very complete characterization of the quantum information theoretic content of Fermi liquids in terms of the geometry of low energy excitations.

\section{One dimensional framework}
The low energy physics of a system of free fermions is given in terms of the geometry of the free Fermi surface.  For example, the heat capacity of such a free fermion system is simply proportional to the density of states times temperature, and the density of states may be written as an integral of $1/v_F$ over the Fermi surface.  Each patch of the Fermi surface contributes to physical quantities like a chiral $1+1$ dimensional free fermion.

Including interactions is possible at the free Fermi fixed point because phase space restrictions reduce the effects of interactions to certain forward scattering terms labeled by an infinite set of Landau parameters.  Each Landau parameter corresponds to a single exactly marginal deformation of the free Fermi fixed point which preserves all scaling dimensions.  These forward scattering interactions preserve the fermion number on each patch of the Fermi surface, and the Fermi liquid has a very large symmetry group: $U(1)^\infty$ \cite{bose1,bose2}.  This one dimensional framework permits computation of the usual physical observables of Fermi liquids, observables like heat capacity and compressibility.  In addition, it provides simple access to many of the anomalous entanglement and fluctuation properties of Fermi liquids.

These anomalous properties characterize the reduced density matrix of a region of linear size $L$ inside the Fermi liquid, and they represent non-traditional global observables as opposed to local correlations functions.  These observables include entanglement entropy, defined as the von Neumann entropy of the reduced density matrix of the region, and fermion number fluctuations, defined as the variance of the fermion density integrated over the region.  The entanglement entropy can be generalized to the R\'{e}nyi entropy defined as
\begin{equation}
S_{\alpha} = \frac{1}{1-\alpha} \ln{\left( \text{tr}(\rho^{\alpha}_L) \right)}.
\end{equation}
The von Neumann entropy $S_{\text{vN}} = - \text{tr}(\rho_L \ln{\rho_L})$ is recovered from $S_{\alpha}$ in the limit $\alpha\rightarrow 1$.  Fermionic systems without a codimension one Fermi surface and most bosonic systems have a R\'{e}nyi entropy which scales with the boundary $L^{d-1}$ of the region considered.  This leading scaling behavior is sensitive to the cutoff of the low energy effective theory and does not define a universal low energy observable.  Other sub-leading terms in the entanglement entropy may provide universal numbers characterizing different phases, but these terms are in general hard to calculate and interpret.

Fluctuations of conserved quantities also behave similarly to the entanglement entropy.  Consider a system of fermions with a conserved fermion number obtained by integrating the density $n(x)$ over the entire system.  Given access to only a subregion of linear size $L$ we may ask about the observable $N_L = \int_{x \in L^d} n(x)$.  The average $\langle N_L \rangle = \text{tr}( \rho_L N_L)$ is simply the average density times the volume of the subregion for a translation invariant system.  The fluctuations $\Delta N_L^2 = \langle (N_L - \langle N_L \rangle)^2\rangle = \text{tr}(\rho_L (N_L - \langle N_L \rangle)^2 )$ are generically non-zero and typically scale as the boundary of the subregion.  However, there are exceptions to this scaling, for example, it receives a logarithmic correction, like the entanglement entropy, for critical one dimensional systems.  Also, symmetry breaking states have fluctuations in conserved quantities that scale as the volume of the subregion; these fluctuations are due to the zero mode of the order parameter.
  
Note that the presence of fluctuations in fermion number within a subregion is not in conflict with a fixed total fermion number for the entire system as we always study small subsystems of a larger system.  Like the entanglement entropy of Fermi liquids, the number fluctuations in a metal with codimension one Fermi surface are anomalously large.  One finds that $\Delta N_A^2 \sim L^{d-1}\ln{L}$ with a multiplicative logarithmic correction \cite{num_fluc,fermion2,fs_fluc_1d,bgs_ferm2}.  We will later determine the precise form of this leading logarithmic term.

These results are not restricted to zero temperature; indeed, there are universal crossover functions that determine how the entanglement entropy and ground state number fluctuations go over into the corresponding thermal quantities as the temperature is raised.  For example, consider a circular region in a two dimensional spherically symmetric Fermi liquid.  The reduced density matrix for this region displays crossover behavior at finite temperature from the zero temperature anomalous entanglement regime to the finite temperature thermal regime.  This crossover is captured by a universal scaling function
\begin{widetext}
\begin{equation}
S(L,T) = \frac{1}{2\pi}\frac{1}{12} \int_k \int_x \, dA_k dA_x |n_x \cdot n_k | \ln{\left(\frac{\beta v_F}{\pi \epsilon} \sinh{\left(\frac{\pi L_{\text{eff}}}{\beta v_F}\right)}\right)},
\end{equation}
\end{widetext}
where $L_{\text{eff}}$ is the chordal distance across the circle.  This crossover function depends only on the geometry of the Fermi surface and the real space region; the physical Fermi velocity $v_F$ appears simply as a unit conversion.  It is the only trace of interactions in this remarkably universal formula.  We will discuss and generalize this formula later in the paper.  A similar formula exists for the number fluctuations which we will derive below.

\section{R\'{e}nyi entropy}
I turn first to the calculation of the R\'{e}nyi entropy of the Fermi liquid.  Like the entanglement entropy, the R\'{e}nyi entropy is a cutoff sensitive quantity.  It typically satisfies a boundary law, but we will find that Fermi liquids have a universal boundary law violating piece tied to the geometry of the Fermi surface.  The R\'{e}nyi entropy is of interest for several reasons.  It is often easier to compute, both in field theory and numerically, for example, by quantum Monte Carlo \cite{renyi1}.  It can also be used to characterize phases of matter, for example in topological phases, conformal field theories, and holographic theories \cite{renyi2,eeqft,renyi3}.  From a quantum information perspective, the notion of single copy entanglement \cite{sce1}, given by the limit $\alpha \rightarrow \infty$ of $S_{\alpha}$, determines the maximal deterministically distillable entanglement from the bipartition of a single copy of a quantum state.  The entanglement entropy only gives the distillable entanglement in the unphysical limit of many copies of the quantum state.  I find that the single copy entanglement is, for large $L$, simply one half the entanglement entropy, similar to the one dimensional conformal result \cite{sce2}.  We will now use the one dimensional framework to compute the R\'{e}nyi entropy of a Fermi liquid.

Restricting first to the non-interacting case, each patch on the Fermi surface is equivalent to the chiral half of one dimensional free fermion.  The R\'{e}nyi entropy $S_{\alpha}$ of such a one dimensional chiral fermion is known from conformal field theory.  In fact, it is simply proportional to the entanglement entropy with an $\alpha$ dependent prefactor.  The precise result is
\begin{equation}
S_{\alpha \, (1+1)} = \frac{1}{2}\left(1 + \frac{1}{\alpha}\right) \frac{c_L + c_R}{6} \ln{\left(\frac{L}{\epsilon}\right)},
\end{equation}
where $L$ is the length of the one dimensional interval and $\epsilon$ is a short distance cutoff.  Returning to the Fermi surface problem, each patch contributes such a term to the entanglement entropy with $c_L = 1$ and $c_R = 0$.  The resulting R\'{e}nyi entropy of the Fermi liquid is remarkably simple; the universal part controlled by the low energy theory is simply
\begin{equation}
S_{\alpha}= \frac{1}{2}\left(1 + \frac{1}{\alpha}\right) S_1,
\end{equation}
where again, $S_1$ is the usual entanglement entropy in two dimensions given by
\begin{equation}
S_1 = S = \frac{1}{2\pi}\frac{1}{12} \int_k \int_x \, dA_k dA_x |n_x \cdot n_k |\ln{\left(\frac{L}{\epsilon}\right)}.
\end{equation}

This isn't the end of the story.  Using the finite temperature crossover form for the von Neumann entropy of the reduced density matrix we obtain a finite temperature crossover form for the R\'{e}nyi entropy.  The result for two dimensions is
\begin{equation}
S_{\alpha} = \frac{1}{2}\left(1 + \frac{1}{\alpha}\right) S(L,T),
\end{equation}
and in any spatial dimension with a codimension one Fermi surface
\begin{widetext}
\begin{equation}
S_{\alpha} = \frac{1}{2}\left(1 + \frac{1}{\alpha}\right) \frac{1}{(2\pi)^{d-1}}\frac{1}{12} \int_k \int_x \, dA_k dA_x |n_x \cdot n_k | \ln{\left(\frac{\beta v_F}{\pi \epsilon} \sinh{\left(\frac{\pi L_{\text{eff}}}{\beta v_F}\right)}\right)}.
\end{equation}
\end{widetext}
I want to point out that since the Fermi surface system has a cutoff, this formula cannot truly be valid for all $\alpha$.  For example, the $\alpha\rightarrow 0$ limit of the R\'{e}nyi entropy always produces the Schmidt rank of the reduced density matrix under study.  However, the $\alpha\rightarrow 0$ limit of the conformal field theory result diverges corresponding to the statement that a theory which is truly conformal to arbitrarily high energy must have an infinite number of local degrees of freedom.  At finite temperature we must keep $T/\alpha < T_F$ or $\alpha > T/T_F$ in order not to probe high energy physics, and similarly, at zero temperature we must keep $\alpha > 1/(k_F L)$ to avoid the influence of non-universal physics away from the Fermi surface.

Now at last restoring the interactions, the complete finite temperature crossover function remains correct even in the presence of interactions because of the nature of the Fermi liquid fixed point (more properly, fixed manifold).  The counting of low energy degrees of freedom remains the same.  There are two possible modifications of this formula when including interactions.  First, the interactions may change the geometry of the low energy Fermi surface.  Spherical symmetry can prevent such a modification, but in a solid state metal the crystal lattice breaks the rotational symmetry.  Nevertheless, the interacting Fermi surface in this less symmetric systems continues to control the R\'{e}nyi entropy.  Second, the Fermi velocity must be replaced by the renormalized physical Fermi velocity.

There are several checks of this formula.  First, it correctly reproduces the one dimensional result that must occur for non-interacting systems with nested Fermi surfaces.  Second, it reproduces the finite temperature R\'{e}nyi entropy of the free Fermi gas; this follows by a direct computation.  Third, it reproduces the thermal R\'{e}nyi entropy of an interacting Fermi liquid.  At finite temperature the R\'{e}nyi entropy is
\begin{equation}
S_{\alpha} = \frac{1}{1-\alpha} \ln{\left(\text{tr}\left(\rho(T)^{\alpha}\right)\right)} = \frac{1}{1-\alpha} \ln{\left(\frac{Z(\alpha \beta)}{Z(\beta)^{\alpha}}\right)}
\end{equation}
where $Z(\beta)$ is the partition function.  The partition function of the Fermi liquid is $- \ln{Z} \sim T$ for small $T$ compared to $E_F$, and this permits us to write $Z(\alpha\beta) = Z(\beta)^{1/\alpha}$.  The final result is that the R\'{e}nyi entropy for an interacting Fermi liquid is the same universal function of $\alpha$ times the thermal entropy as predicted by the formula above.

\section{Entanglement entropy of disjoint regions}
Having described in detail the leading behavior of the R\'{e}nyi entropy for a single convex region, I now turn to the problem of describing the entanglement entropy for more complicated subregion geometries.  Perhaps the most important motivation for this study is the calculation of the quantum mutual information $I(A,B)$ between two regions $A$ and $B$.  The mutual information is
\begin{equation}
I(A,B) = S_A + S_B - S_{A \cup B},
\end{equation}
and this definition requires knowledge of the entanglement entropy of disjoint regions to compute.  We will see that the entropy of disjoint regions can be computed in a manner similar to that of convex regions via a more general formula that can handle arbitrary region geometry.  The basic strategy remains the same: express all quantities in terms of sums over one dimensional modes on the Fermi surface.  Because these one dimensional modes are conformal and, in fact, essentially free fermions, much can be calculated.  In particular, the main information we need is the entanglement entropy of a one dimensional conformal field theory when the one dimensional subregion consists of multiple regions.

This quantity, the entanglement entropy of multiple regions in a conformal field theory, is not a simple quantity to compute in general.  Despite some early claims in the literature it is not as universal as the single region entanglement entropy which depends only on the central charge of the conformal field theory. The multi-interval entanglement entropy depends on the entire operator content of the theory as it requires evaluating higher point correlation functions of twist fields.  However, the relative simplicity of the Fermi liquid is a boon: the multiple interval result for a one dimensional free fermion is known.  We will see shortly how this information permits calculation of the entanglement entropy of arbitrary regions in a Fermi liquid.

Let us briefly recall the one dimensional result.  We consider a subsystem consisting of several disjoint intervals labeled $[a_i, b_i]$  $i=1, ...,m$ with $m$ the number of intervals.  As we said before, the entanglement entropy of such a subregion is in general complicated, but for free fermions the answer is remarkably simple.  The result is
\begin{widetext}
\begin{equation}
\label{1dmi}
S_{1+1} = \frac{c_L + c_R}{6}\left( \sum_{ij} \ln{\left|a_i - b_j \right|} - \sum_{i < j} \ln{\left|a_i - a_j \right|} - \sum_{i < j} \ln{\left|b_i - b_j \right|}\right),
\end{equation}
\end{widetext}
with the implicit presence of cutoffs in the logarithms is understood.  This formula reduces to the usual result in $1$ dimension in the case of a single interval.  Note that when computing the mutual information, the number of logs being added and subtracted is equal giving a result independent of the cutoff.

\begin{figure}
\includegraphics[width=.5\textwidth]{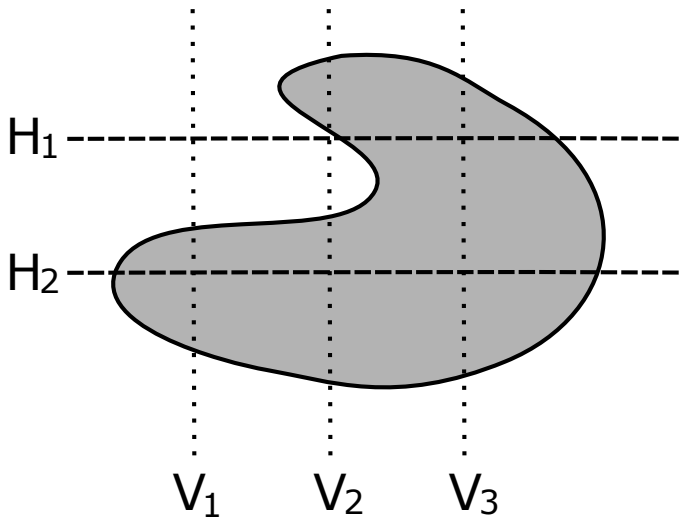}
\caption{Geometry of the entanglement entropy for a non-convex real space region.  The shaded grey region surrounded by the thick black line represents a particular non-convex real space region.  Dotted lines labeled $V_i$ ($i=1,2,3$) represent particular one dimensional cuts experienced by a patch of the Fermi surface with vertical Fermi velocity.  Similarly, dashed lines labeled $H_i$ ($i=1,2$) represent particular one dimensional cuts experienced by a patch of the Fermi surface with horizontal Fermi velocity.  Notice how the vertically moving fermion switches from an effectively single interval geometry in $V_1$ to a two interval geometry in $V_2$ and back to a single interval in $V_3$.  On the other hand, $H_1$ and $H_2$ are both effectively single interval geometries. }
\end{figure}

How do we use this formula to compute the entanglement entropy of a Fermi liquid in an arbitrary geometry?  The procedure is illustrated in Fig. 1 and goes as follows.  First, we choose a point on the Fermi surface; this point defines a direction in space via the local Fermi velocity.  Second, we choose a point on the real space boundary and draw a line through this point with direction determined by the Fermi velocity from step one.  Third, we determine the intersections of this line with the boundary of the real space region; these intersection points are grouped into effectively one dimensional intervals.  The contribution to the entropy of such a configuration is given by a slightly modified version of the one dimensional result above (\ref{1dmi}).  To each logarithm in the one dimensional result we append geometrical ``flux factors" $|n_x \cdot n_k |$ times a differential area element $dA_x$ where $n_k$ is the direction determined by the Fermi velocity and $n_x$ is a real space normal vector.  In detail, each logarithmic term of the form $\ln{|x-y|}$ is multiplied by a factor of $\frac{|n_x \cdot n_k | dA_x + |n_y \cdot n_k| dA_y}{2}$, where $n_x$ and $n_y$ are the local real space boundary normals.  As illustrated in Fig. 2, the flux factors and differential area elements satisfy $|n_x \cdot n_k | dA_x = |n_y \cdot n_k | dA_y$.  Finally, the contribution of this configuration of intervals must be divided by the number intersection points to avoid over-counting and integrated over all such points on the real space boundary and the Fermi surface.  This geometrical calculation gives the leading boundary law violating contribution to the entanglement entropy of a Fermi liquid for arbitrary Fermi surface shape and real space geometry.

This formula is well defined but complex. There are a number of simple checks that can be performed.  For two or more well separated convex regions, the formula above reduces to a sum of terms corresponding to the entanglement entropy of each region taken separately.  For free fermions with a nested Fermi surface one immediately recovers the one dimensional result as all the flux factors become trivial.  At finite temperature, all the logarithms $\ln{L}$ are replaced by crossover functions of the form $\ln{\left(\sinh{\left(\pi L T/ v\right)}\right)}$ leading to a simple form at high temperature that is extensive in the total real space region size.  This one dimensional crossover form is well known for single intervals in one dimensional conformal field theory, but it is nontrivial for the multiple interval case.  It comes from evaluating higher point correlations functions of twist fields on a Euclidean torus in the free fermion conformal field theory.  As before, the simple form depends crucially on the fact that we are dealing with free fermions; this result does not hold in a general conformal field theory.

The absence of Landau parameters is reasonable given the fact that the entanglement entropy (at least for a single interval) doesn't depend on interaction strength even for a Luttinger liquid in one dimension.  I cannot completely rule out the possibility that there is a prefactor $f$ in the crossover function which is a function only of $LT$: $f = f(LT)$ with the limits $f(x\rightarrow \infty) \rightarrow 1$ and $f(x\rightarrow 0) \rightarrow \text{constant}$, but such a prefactor seems unlikely.  Given the absence of Landau parameters at high temperature and the large $U(1)^{\infty}$ symmetry, I believe the Fermi liquid entanglement entropy is universal as I have described.

\begin{figure}
\includegraphics[width=.5\textwidth]{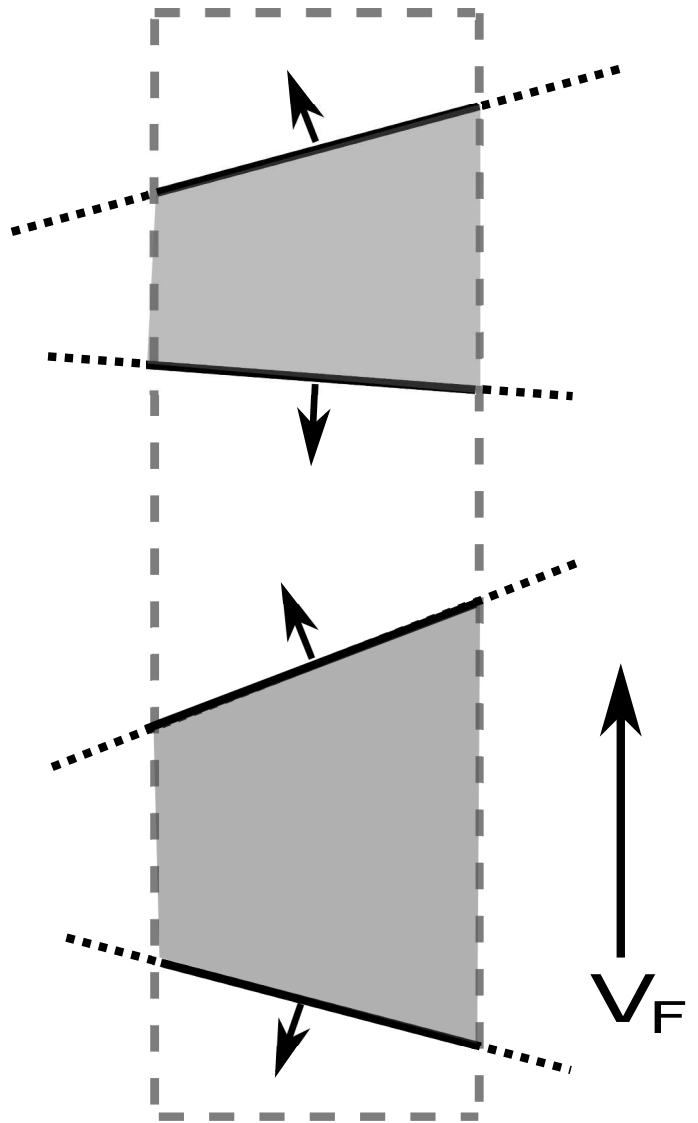}
\caption{Multiple interval geometry and flux factors.  A sequence of intervals similar to the situation shown in Fig. 1 $V_2$ with the local Fermi velocity vertical.  The dashed grey line encloses the real space boundary segments of interest; the segments themselves are the black lines which continue into dotted lines outside the dashed grey enclosing line.  The dark grey shaded regions are interior regions of the real space region, while arrows at the surfaces of the these regions indicate local normals to the real space boundary.}
\end{figure}

There are also some subtleties in the evaluation of this formula.  The most basic subtlety is that the number of intervals is subject to change depending on the point on the real space boundary chosen.  These weak singularities are nevertheless completely harmless and do not render the integral ill defined.  It is best to illustrate the formula by a simple example; further results of this nature will be presented elsewhere.  Take a $d$ dimensional spherically symmetric Fermi liquid, and consider a real space region consisting of two spheres $A_1$ and $A_2$ of radius $R$ separated by a center-to-center distance $\ell$.  A very interesting quantity is the mutual information between the two spheres defined as $I(R,\ell) = S_{A_1} + S_{A_2} - S_{A_1 \cup A_2}$.  This quantity is quite powerful as it bounds normalized connected correlation functions between operators localized in the two spheres.  We now evaluate this quantity for the two sphere geometry in the limit $\ell \gg R$ thus bounding the long distance decay of correlations in the Fermi liquid.

Because we subtract the entropy of each region separately, the only contribution to the mutual information comes from configurations that involve both spheres.  As shown in Fig. 2, each pair consisting of a Fermi surface point and a point on the real space boundary only contributes if the line drawn from the real space point parallel to the local Fermi velocity intersects both spheres.  The fraction of pairs of points satisfying this constraint vanishes in the limit $\ell/R \rightarrow \infty$ which is the statement that the mutual information vanishes for infinite sphere separation.  Our task is to estimate how this fraction vanishes in said limit.  Let one sphere sit at $z=0$ and the other sit at $z = \ell$; both spheres sit at $x_i = 0$ for $i=1...d-1$ with $x_d = z$.  Since I have assumed spherical symmetry, the Fermi surface is sphere of radius $k_F$ in momentum space with $k_F$ fixed by density.  Now the Fermi surface points with $\vec{v}_F = \pm v_F \hat{z}$ always gives lines connecting the two spheres, but as $\ell/R \rightarrow \infty$, only a small neighborhood around these points continue to generate lines connecting the two spheres.  We can estimate the linear size $\delta k$ of this neighborhood as $\delta k \sim k_F \left(\frac{\ell}{R}\right)$.  For $\ell/R \rightarrow \infty$, the difference of logarithms appearing the mutual information nearly cancels with a residual $(R/\ell)^2$ falloff.  Putting everything together, I find that roughly $(R \delta k)^{d-1} $ patches on the Fermi surface each contribute $(R/\ell)^2$ to the mutual information for a total mutual information going like
\begin{equation}
I(\ell,R) \sim (R k_F)^{d-1} \left(\frac{R}{\ell}\right)^{d+1}.
\end{equation}

\begin{figure}
\includegraphics[width=.5\textwidth]{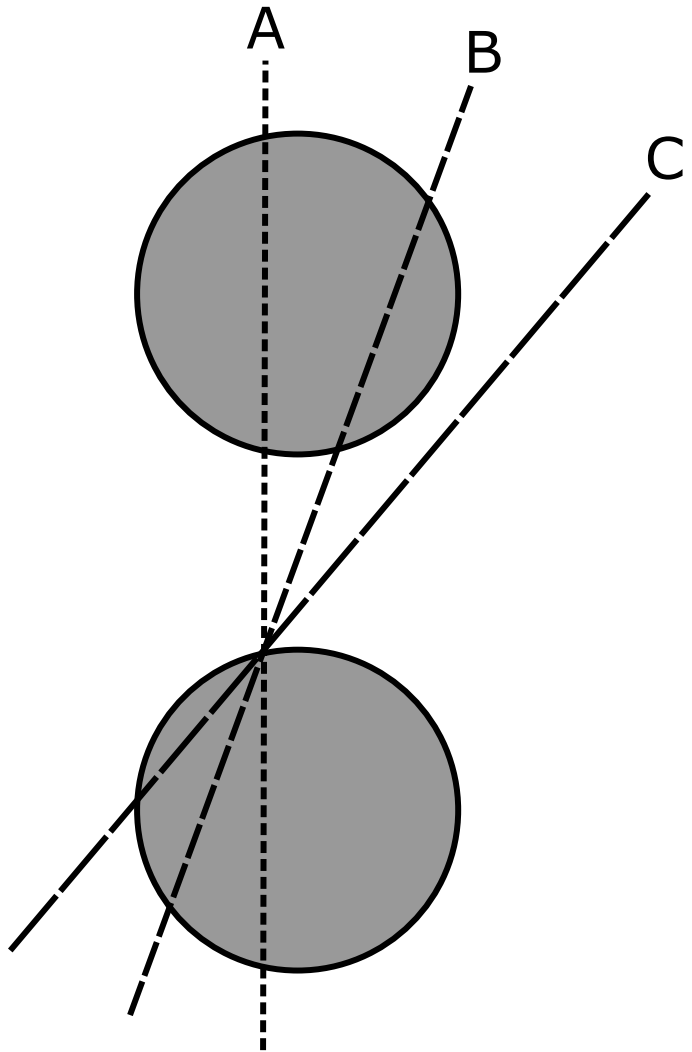}
\caption{Geometry of the mutual information for disconnected real space regions.  The three one dimensional cuts labeled A, B, and C represent three different patches on the Fermi surface for different angles of the Fermi velocity.  Cut A comes from a mode with vertical Fermi velocity; it always cuts both spheres and hence always contributes to the mutual information in this geometry.  Cut B also connects both spheres, at least for some choices of real space boundary point.  However, cut C never intersects both spheres simultaneously and hence does not contribute to the mutual information.  For the sphere geometry shown here, there is always a maximum angle from vertical such that cuts beyond that angle never intersect both spheres.  As described in the text, this maximum angle approaches zero as the spheres are taken far apart relative to their size.}
\end{figure}

As I have already indicated, the mutual information provides a bound on connected correlation functions.  The precise statement is that the square of the connected correlation function of any two operators, normalized by the operator norms, is bounded by the mutual information.  Thus we learn that connected correlation functions of local operators in a Fermi liquid must decay at least as $ x^{-\frac{d+1}{2}}$ as $x \rightarrow \infty$.  Calculations are trivially possible for free fermions, and one can directly verify that all two point functions of bosonic composite operators do indeed decay fast enough to satisfy the bound.  The bound is saturated by the two point function of fermions which decays as $ x^{-\frac{d+1}{2}}$ times an oscillating function of $k_F x$.  This is not completely trivial since the fermion creation operator is not strictly speaking a local operator i.e. in the Jordan-Wigner representation, but this mild non-locality does not to spoil the utility of the mutual information.  The exact coefficient is calculable by performing the geometrical integrals outlined above.  On the other hand, nested Fermi surfaces give a mutual information that only decays as $1/\ell^2$ indicating the possible presence of slowly decaying correlations.  Indeed, this bound is again saturated by the fermion two point function which decays as $1/x$ for a nested Fermi surface with $x$ parallel to the nesting vector.

I have also numerically computed the mutual information between two rectangular blocks in a two dimensional free fermion system with a nearly spherical Fermi surface. The geometrical integrals can be done in the limit $R \gg W$ with the result
\begin{equation}
I \sim \frac{k_F R}{6 \pi} \frac{W^2 R}{L^3}
\end{equation}
where $W$ is the box width, $R$ is the box length, and $L$ is the separation between the boxes.  This expression agrees with the numerical result to within a few percent with the deviation due mostly to corrections in $W/R$ and $R/L$; additionally, the qualitative $1/L^3$ dependence also agrees with the numerical results.

\section{Number fluctuations}
Finally, I turn to a description of the number fluctuations.  Here at last Fermi liquid parameters will make an appearance.  Recall that the basic result on the number fluctuations is that they scale as $L^{d-1} \ln{L}$ in a Fermi liquid for a real space region of linear size $L$.  For free fermions the precise prefactor is known both from operator methods and from the conformal field theory approach advocated here.  Unlike the result for entanglement entropy, the number fluctuations do depend on Landau parameters as interactions are turned on.  To see this note that the same connected correlation function used to determine the number fluctuations is also related to the compressibility at finite temperature.  This compressibility does depend on a Landau parameter in addition to the effective mass through the density of states.  Thus we can already say that the number fluctuations $\Delta N_L^2(L,T)$ must depend on the Landau parameters in the limit $T \rightarrow \infty$ (always keeping $T \ll E_F$.)  In fact, we can say much more because in Fermi liquid theory the interacting density-density correlation function is related to the free correlation function by a series of forward scatting diagrams that can be summed to give $1/(1+F_0)$ where $F_0$ is the familiar $\ell =0 $ Landau parameter for spin-less fermions.  The number fluctuations in a $d$ dimensional Fermi liquid at zero temperature take the asymptotic form
\begin{equation}
\Delta N_L^2 \sim \frac{1}{1+F_0}\frac{1}{(2\pi)^{d-1}}\frac{\ln{L}}{4 \pi^2}\int_x \int_k dA_x dA_k |n_x \cdot n_k|,
\end{equation}
with $L$ the linear size of the convex region considered.

In the presence of interactions at finite temperature I find the familiar crossover form
\begin{widetext}
\begin{equation}
\Delta N_L^2(L,T) =  \frac{1}{1+F_0}\frac{1}{(2\pi)^{d-1}}\frac{1}{4 \pi^2} \int_k \int_x \, dA_k dA_x |n_x \cdot n_k | \ln{\left(\frac{\beta v_F}{\pi \epsilon} \sinh{\left(\frac{\pi L_{\text{eff}}}{\beta v_F}\right)}\right)},
\end{equation}
\end{widetext}
where the $\ln{L}$ term has been replaced by its finite $T$ generalization $\ln{\left(\sinh{\left(\pi L T /v \right)}\right)}$.  As before, one can check that in the limit of high temperature one recovers the usual number fluctuations proportional to region size and depending on $F_0$ and the physical Fermi velocity $v_F$.

The generalization to non-convex regions or multiple regions is also possible in line with the generalization of the entanglement entropy discussed above.  Indeed, the structure of logarithms in (\ref{1dmi}) also appears for the number fluctuations in a one dimensional Fermi liquid.  The density-density correlation function is
\begin{equation}
\langle n(x) n(y) \rangle_c \sim \frac{1}{(x-y)^2},
\end{equation}
and upon integrating $x$ and $y$ over different intervals I find exactly the structure of sums of logarithms appearing in (\ref{1dmi}).  This means that the same geometrical integral that gave the multiregion entanglement entropy in the higher dimensional Fermi liquid also appears in the calculation of the number fluctuations.  Indeed, these two quantities only differ in the prefactor of the geometrical integral describing the interplay of real space and Fermi surface geometries.  To be precise, in the limit of large $L$ with a fixed high energy cutoff, I find that $\Delta N_L^2 /S_L = \frac{3}{ \pi^2} \frac{1}{1+F_0}$.  In fact, given the essentially identical forms of the postulated finite temperature scaling functions for entropy and number fluctuations, this ratio is actually valid even at finite temperature.

\section{Discussion}
I have described the calculation of R\'{e}nyi entropy, mutual information, and number fluctuations in Fermi liquids in any dimension for arbitrary codimension one Fermi surface and sub-region geometry.  These results give a very complete description of the low energy structure of quantum information in a Fermi liquid.  This basic Fermi liquid state which underlies so many materials is a truly highly entangled quantum phase of matter.  Remarkably, the entire structure of quantum information in these systems is controlled by a beautiful interplay between the geometry of the Fermi surface and the structure of one dimensional conformal field theory.

There is one subtlety which deserves a more careful discussion, namely the issue of singularities on the Fermi surface.  The methods described above apply directly to smooth Fermi surfaces, but the formalism is strong enough to handle aspects of singular situations as well.  To give a simple example, consider spinless fermions hopping on a square lattice at half filling.  The Fermi surface consists of four straight lines running diagonally between the midpoints of the Brillouin zone boundary.  There are singularities on the Fermi surface at $(\pm \pi, 0)$ and $(0, \pm \pi)$, points where the Fermi velocity vanishes leading to well known singularities in the density of states.  As an example of the subtleties that may arise, consider the mutual information between two regions separated by a large distance compared to their size along the y-axis.  A little thought reveals that the leading contribution to the mutual information from the Fermi surface, in the form presented above, vanishes in this geometry.  In fact, the mutual information is not totally zero; indeed, it cannot be since the fermion correlation function doesn't vanish in the limit of large y.  However, the fermion correlation function does decay faster ($1/y^2$) than the result for a non-singular Fermi surface ($1/y^{3/2}$).  In this sense, the vanishing of the leading ``Fermi surface" part of the mutual information is physical, telling us that there are no extremely long ranged correlations such as would come from a non-singular Fermi surface.  When the Fermi surface geometry does lead to extremely long ranged correlations, such as along the nesting vector $(\pi, \pi)$ in the half filled example, then the mutual information formalism above perfectly captures the requisite decay.  The simple lesson is that there are always subleading terms in all quantities which can become visible if the Fermi surface contribution happens to vanish.  We must simply always remember that the formalism developed in this paper captures only the leading Fermi surface contribution to the physics.

It should be clear by now that there are other quantities of interest that can be calculated for Fermi liquids using the one dimensional approach.  Particularly interesting would be to study the behavior of entropy and other quantities in a non-equilibrium situation, following a quench of some type, for example.  Powerful techniques from one dimensional conformal field should give a great deal of control over the non-equilibrium structure of entanglement in Fermi liquids.  There is still the question of experimental observation of the effects described here.  The number fluctuations are in principle easy to observe, we simply need to prepare a clean Fermi liquid, perhaps in an optical lattice, and repeatedly count the number of fermions in a given subsystem to evaluate $\Delta N^2$.  If the subregion can be made large enough I have a quite precise prediction for $\Delta N^2$.  Another route is to actually physically affect the separation of a given subregion from a larger system and measure the resulting entropy thermodynamically.  I am in the process of studying the details of this process.

Given their experimental ubiquity, it is gratifying to have control of the quantum information content of Fermi liquids.  Of course, in the case of Fermi liquids in the solid state, the logarithms will be cutoff by the mean free path in the presence of disorder.  Still, any clean Fermi liquid system with a long mean free path will display the anomalous quantum information theoretic properties described here.  In the larger framework of many-body entanglement, Fermi liquids have taught us how to produce one class of highly entangled quantum states.  However, much remains to be understood about the role of entanglement in deeply quantum mechanical phases, especially gapless phases like those explored here \cite{nfl1,crit_fs,fs_boson1,fs_boson2,dbl}.  Like Fermi liquids, these phases should violate the boundary law for entanglement entropy.  And the search continues for other highly entangled phases of quantum matter.

\section{Acknowledgements}
I thank Xiao-Gang Wen for support and Matt Hastings for discussions about fluctuations.  I also thank the 2010 Boulder School for support during the final stages of this work.

\bibliography{renyi_ff} 
\end{document}